\documentclass[prb,aps,twocolumn,superscriptaddress,showpacs]{revtex4}
\usepackage{graphicx}
\usepackage{psfrag}
\usepackage{color}
\usepackage{subfigure}
\usepackage{amsmath}
\definecolor{dred}{rgb}{0.7,0.0,0.0}
\allowdisplaybreaks
  
\begin{document}

%
%

\title{Role of Degeneracy, Hybridization, and Nesting \\
in the Properties of Multi-Orbital Systems}

\author{Andrew Nicholson}
\author{Qinlong Luo}
\author{Weihao Ge}
 
\affiliation{Department of Physics and Astronomy, The University of
  Tennessee, Knoxville, TN 37996}
\affiliation{Materials Science and Technology Division, Oak Ridge
  National Laboratory, Oak Ridge, TN 32831}
 
\author{Jos\'e Riera}
 
\affiliation{Instituto de F\'isica Rosario, Universidad Nacional de Rosario, 2000-Rosario, Argentina}
 

\author{Maria Daghofer}
 
\affiliation{IFW Dresden, P.O. Box 27 01 16, D-01171 Dresden, Germany}

\author{George B. Martins}
 
\affiliation{Department of Physics, Oakland University, Rochester,
Michigan 48309}

\author{Adriana Moreo}
\author{Elbio Dagotto}

\affiliation{Department of Physics and Astronomy, The University of
  Tennessee, Knoxville, TN 37996}
\affiliation{Materials Science and Technology Division, Oak Ridge
  National Laboratory, Oak Ridge, TN 32831}

\date{\today}
 
\begin{abstract}
 
To understand the role that degeneracy, hybridization, and nesting
play in the magnetic and pairing properties of multiorbital Hubbard models 
we here study numerically two types of two-orbital models, both with hole-like and 
electron-like Fermi surfaces (FS's) that are related by nesting vectors $(\pi,0)$ and $(0,\pi)$. In
one case the bands that determine the FS's arise from strongly hybridized
degenerate d$_{xz}$ and d$_{yz}$ orbitals, while in the other the two bands are
determined by non-degenerate and non-hybridized $s$-like orbitals. Using a variety
of techniques, in the weak coupling regime it is shown that only the model with 
hybridized bands develops metallic magnetic order, while the other model exhibits an
ordered excitonic orbital-transverse spin state that is insulating and does not have a local magnetization. 
However, both models display similar insulating magnetic stripe ordering in the strong
coupling limit. These results indicate that nesting is a necessary but not sufficient condition 
for the development of ordered states with finite local magnetization in multiorbital Hubbard systems; 
the additional ingredient appears to be that the nested portions of the bands need to have the same orbital flavor. 
This condition can be 
achieved via strong hybridization of the orbitals in weak coupling or via the FS reconstruction induced 
by the Coulomb interactions in the strong coupling regime. This effect also impacts the pairing 
symmetry as demonstrated by the study of the dominant pairing channels for the two models.
 
 
\end{abstract}

\maketitle

\section{Introduction}
 
 
Among the several aspects of the study of the
iron-based superconductors that are still controversial 
and unsettled,~\cite{johnston} the following two questions have attracted 
considerable attention: {\it (i)} Does the magnetic order
observed in the parent compounds\cite{dai} arise from the nesting
properties of the non-interacting (or high temperature) 
Fermi surface\cite{mazin,kuroki} or should a better description be based
on the superexchange Heisenberg
interactions between localized magnetic moments?\cite{si} {\it (ii)} What is
the pairing mechanism, to what extent is the pairing symmetry determined by nesting and, what is
the actual symmetry and momentum dependence 
of the pairing operator? In particular, what is 
the role that the orbital degrees of freedom play in this context? 

The origin of the magnetic state is being vigorously debated. One proposal, 
based on fermiology, is the excitonic mechanism in which electron-hole pairs are formed
by one electron and one hole from different FS's nested with nesting vector 
${\bf Q}$. In this context some studies disregard the orbital structure of the 
bands\cite{mazin,chubu,tesa,phil} while others stress the role played by their orbital
composition.\cite{raghu,Daghofer:2009p1970,graser,moreo,kemper,rong,our3}  
Another approach focuses on the order of the localized moments that develop in the presence of
strong Coulomb interactions\cite{si,yildirim,uhrig,kruger} and relies on ab initio 
results\cite{nakamura,anisimov} that suggest that 
the pnictides are moderately, rather than weakly, correlated, conclusion supported by photoemission 
measurements
indicating mass enhancements due to electron correlations as large as 2-3.\cite{lu} 

The pairing mechanism in the pnictides is also controversial. Most of the pairing operators that have been
proposed in the literature either ignore the multi-orbital characteristics
of the problem or consider
Cooper pairs that are made out of electrons located at the same orbital. 
A majority of these  previous studies have been performed in the weak coupling limit. 
The original proposal of the $s_{\pm}$ pairing state
dealt with the overall symmetry of the pairing operator but without
distinguishing among the spatial vs. orbital contributions to its particular form.\cite{mazin,kuroki}
Other authors~\cite{schmalian} have considered a
spin-fluctuation-induced pairing interaction and also assumed that Cooper pairs
are predominantly made of electrons in the same orbital.
A Random-Phase Approximation (RPA) analysis~\cite{graser} concluded
that the pairing is, again, intraorbital, both for the A$_{1g}$ ($s$-wave) and B$_{1g}$ ($d$-wave)
symmetries. Among the authors that have used the conceptually different 
strong coupling approach, some have
studied effective single orbital models~\cite{si} while others incorporated two orbitals,~\cite{goswami} but still only considering intra-orbital
pairing operators. The same model was also studied under a mean-field approximation~\cite{seo} 
with the assumption
that exchange takes place between spins on the same orbitals and, again,
only intraorbital pairs were proposed.
 
Among the early first studies of multiband superconductors, 
Suhl {\it et al.}\cite{suhl} considered two tight-binding bands, hypothetically
identified with $s$ and $d$ orbitals, and the effect of weak electron-phonon
interactions. Under these assumptions, it was reasonable to expect that the
Cooper pairs would be formed by electrons belonging to the same band. However, the
actual orbital composition of the pairs was not addressed. The interacting
portion of the Hamiltonian was written in the band representation and this model
was proposed by analogy with models used in the BCS theory, 
assuming that emission and absorption
of a phonon could occur in four ways. These four processes corresponded
to pair scattering within each of the two FS's and pair hopping from 
one FS to the other.
This last process would occur if the exchanged phonon has enough momentum to
allow the Cooper pair to jump from a FS to the other, and it can occur even if
the orbitals do not hybridize to form the bands.\cite{hotta} In this case,
the expected pairing operator is the traditional on-site $s$-wave state 
of the BCS theory, with a momentum independent gap. 
In principle, independent gaps may arise
on the different FS's\cite{suhl} unless the orbitals are hybridized by
the symmetries of the Hamiltonian, in which case the gaps
will have to be related to each other and obey the symmetries of
the system.\cite{moreo} 

The previous discussion applies to superconductors driven by the electron-phonon
interaction. However, it is believed that the most relevant interactions in the pnictides are 
the Coulomb repulsion and Hund magnetic exchange. These interactions are more easily expressed in
real space and in the orbital representation. In fact, 
the effective form of the Coulomb interaction in the band representation is
more complicated than the expression provided by Suhl {\it et al.}\cite{suhl} for
the electron-phonon interaction. In particular, it has been shown~\cite{chubu} 
that a pair hopping term, such as the one
introduced by Suhl {\it et al.} occurs only if the orbitals get hybridized to form
the bands. If the orbitals are $not$ hybridized this type of
term is not present in the effective interaction Hamiltonian.
In addition, when the bands are made of hybridized orbitals, as it is
the case for the iron
pnictides,\cite{boeri} the actual orbital structure of the pairs needs to be
considered since due to the Coulomb repulsion on-site pairing is not expected
to occur, and the overall symmetry properties of the pairing operators may be
a function of their spatial and orbital
components.~\cite{Daghofer:2009p1970,moreo}
 
To understand the role that the orbitals play in the case of electrons with
strongly hybridized bands that are interacting via the Coulomb repulsion, 
as believed to occur in the case of
the pnictides in the context of the magnetic scenario for superconductivity, 
in this manuscript we present and discuss Lanczos numerical, Hartree mean-field, and
RPA studies of 
two different two-orbital models, both displaying identical Fermi surfaces.
One of them is the well-known and widely used
two-orbital model for the pnictides\cite{raghu,Daghofer:2009p1970,moreo} based
on the two strongly hybridized degenerate d$_{xz}$ and d$_{yz}$ orbitals of
iron, while the second is a two-band ``toy- model'' (dubbed the $s$-model) whose bands arise from two
non-hybridized, non-degenerate, $s$-like orbitals that is introduced here for
the first time. The latter model has a FS
qualitatively similar to that of the pnictides. In both cases a hole
(electron) FS is located at the $\Gamma/M$ ($X/Y$) points of the Brillouin zone
(BZ). The hole and electron FS's are connected by nesting vectors $(\pi,0)$ and $(0,\pi)$. The
role that the nesting and the orbitals play in the magnetic and pairing
properties of these models will be here investigated and discussed, both in the weak
and strong coupling regimes.
       
Besides its conceptual relevance, the results presented here should also be
framed in the context of recent bulk-sensitive laser angle-resolved photoemission (ARPES)
experiments~\cite{shimojima} on BaFe$_2$(As$_{0.65}$P$_{0.35}$)$_2$ and 
Ba$_{0.6}$K$_{0.4}$Fe$_2$As$_2$. The main conclusion of Ref.~\onlinecite{shimojima}
is the existence of orbital independent superconducting gaps that are not expected
from spin fluctuations and nesting mechanisms, but are claimed 
to be better explained by magnetism-induced
interorbital pairing and/or orbital fluctuations. 
This is argued based on the observation that
the $3z^2 − r^2$ orbital that
forms one of the hole pockets at the BZ center, but that
does not have a nested partner with the same orbital at
the electron pockets, nevertheless appears to develop a
superconducting gap.
Another interesting experimental result that challenges the role of nesting in the physics
of the pnictides is a careful measurement of the de Haas-van Alphen (dHvA) effect in BaFe$_2$P$_2$, the end member of the series 
BaFe$_2$(As$_{1-x}$P$_x$)$_2$, indicating that this non-magnetic and non-superconducting 
compound displays the best nesting
of all the compounds in the series.\cite{arnold}

The manuscript is organized as follows. In Section~\ref{model} the models are
introduced. The magnetic properties are presented in Section~\ref{magnetic}
while the pairing properties are the subject of Section~\ref{pairing}.
Section~\ref{conclusions} is devoted to the conclusions.

\section{Models}\label{model}
 
\subsection{$d$-model}\label{dmodel}
The reference model that will be considered here is the widely-used
two-orbital model~\cite{Daghofer:2009p1970,moreo,raghu} based on the
d$_{xz}$ ($x$) and d$_{yz}$ ($y$) Fe orbitals of the pnictides.
Since the two orbitals are degenerate, 
an important detail is that the
direction along which each orbital is defined is actually arbitrary. Two directions have
been used in the literature: $x,y$\cite{raghu,Daghofer:2009p1970,moreo} with
the $x$ and $y$ axes along the directions that connect nearest-neighbor iron
atoms, and $X,Y$\cite{kuroki,dhlee} with the $X$ and $Y$ axis rotated 45$^o$
with respect to the $(x,y)$ set. In terms of the d$_{xz}$ and d$_{yz}$ orbitals
the tight-binding dispersion of the two-orbital model is given by\cite{wan}
 
\begin{equation}
\begin{split}
\xi_{xy}({\bf k})=[-(t_1+t_2)(\cos k_x + \cos k_y)-\\
4t_3 \cos k_x \cos k_y-
\mu]\tau_0\\
-(t_1-t_2)(\cos k_x - \cos k_y)\tau_3-\\
4 t_4\sin k_x \sin k_y \tau_1,
\end{split}
\label{xydis}
\end{equation}
\noindent where $\tau_i$, with $i=1,2,3$, are the Pauli matrices and $\tau_0$ is
the $2 \times 2$ identity matrix. The $\tau_i$ matrices act in orbital space. 
Note that $\xi_{xy}({\bf k})$ must
transform as the A$_{1g}$ representation of D$_{4h}$; in this representation
$\tau_0$ transforms as A$_{1g}$, $\tau_1$ as B$_{2g}$, and $\tau_3$ as
B$_{1g}$. However, if the degenerate $d$-orbitals are expressed 
in terms of the $(X,Y)$ axes as
$(d_X,d_Y)$, then the tight-binding dispersion becomes:\cite{dhlee}
\begin{equation}
\begin{split}
\xi_{XY}({\bf k})=[-(t_1+t_2)(\cos k_x + \cos k_y)-\\
4t_3 \cos k_x \cos k_y-
\mu]\tau_0\\
-(t_1-t_2)(\cos k_x - \cos k_y)\tau_1-\\
4 t_4\sin k_x \sin k_y \tau_3.
\end{split}
\label{XYdis}
\end{equation}
\noindent
 
\noindent Notice that the $XY$ basis is chosen just for the orbitals while in real space
the system of coordinates is still given by $(x,y)$. $\xi_{XY}({\bf k})$ also has to transform 
as A$_{1g}$ which means that, when the
orbitals are defined in the $(X,Y)$ basis, then $\tau_1$ transforms as B$_{1g}$,
and $\tau_3$ as B$_{2g}$. It can be shown that since $\tau_1$ is the matrix
that indicates inter-orbital electron hopping, this kind of hopping happens
between nearest-neighbors [next-nearest-neighbors] in the $(X,Y)$ [($x,y$)]
representation.
 
\begin{figure}[thbp]
\begin{center}
\includegraphics[width=8.0cm,clip,angle=0]{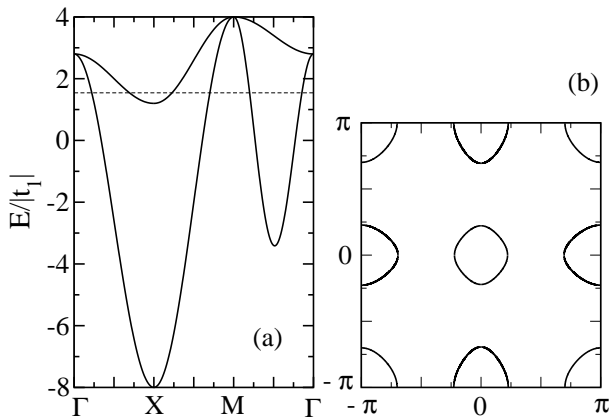}
\caption{(a) Band dispersion and (b) Fermi surface for
the half-filled two-orbital $d$-model for the pnictides, with the hopping
parameters introduced in Ref.~\onlinecite{raghu}.}
\vskip -0.5cm
\label{dorbital}
\end{center}
\end{figure}

As previously discussed, if the values of the parameters are set to $t_1=-1$, $t_2=1.3$,
$t_3=t_4=-0.85$ and $\mu=1.54$ then the Fermi surface (shown in
Fig.~\ref{dorbital} together with the band dispersion) for the
tight-binding Hamiltonian is in qualitative agreement with band structure
calculations for the pnictides~\cite{raghu} once folding to the reduced BZ is
performed. Note that the system is half-filled (two electrons per Fe site on average) 
and, due to the orbital degeneracy, each orbital is half-filled as well, despite the fact that
the bands are not equally filled.
 
An important characteristic of the two degenerate $d$-orbitals in this model is
that around the hole pockets a spinor describing the mixture of orbitals
rotates twice on encircling these FS's. The inversion and time reversal
symmetry of the
twice degenerate $d$ bands ensures that at each ${\bf k}$ point it is
possible to choose real spinor wavefunctions that are confined to a
plane. The spinor has vorticity $\pm 2$ around the hole pockets
while there is no vorticity around the electron pockets.\cite{dhlee}
As pointed out in Ref.~\onlinecite{dhlee}, this topological characterization of
the hole and electron pockets is also a characteristic of all the more realistic models
for the pnictides that include additional orbitals.
 
\subsection{$s$-model}\label{smodel}
 
Let us introduce now a two-orbital model with two non-degenerate
non-hybridized $s$-like bands, called $s_1 (1)$ and $s_2(2)$, with dispersion relations
given by:
\begin{equation}
\xi_{s_1}({\bf k})=2t_1(\cos k_x + \cos k_y)+4t_2 \cos k_x \cos k_y-
\mu,
\label{s1dis}
\end{equation}
\noindent and
\begin{equation}
\xi_{s_2}({\bf k})=2t_3(\cos k_x + \cos k_y)+4t_4 \cos k_x \cos k_y-
\mu+\Delta,
\label{s2dis}
\end{equation}
\noindent where $\mu$ is the chemical potential and $\Delta$ is the energy
difference between the two bands. The dispersions can also be written in the basis $(s_1,s_2)$, i.e., $(1,2)$, using the
$\tau_i$ matrices as in the previous case:
 
\begin{equation}
\begin{split}
\xi_{S}({\bf k})=[(t_1+t_3)(\cos k_x + \cos k_y)+\\
2(t_2+t_4) \cos k_x \cos k_y-
\mu+{\Delta\over{2}}]\tau_0\\
+[(t_1-t_3)(\cos k_x + \cos k_y)+\\
2 (t_2-t_4)\cos k_x \cos k_y-{\Delta\over{2}}]\tau_3.
\end{split}
\label{Sdis}
\end{equation}
\noindent
 
\begin{figure}[thbp]
\begin{center}
\includegraphics[width=8.0cm,clip,angle=0]{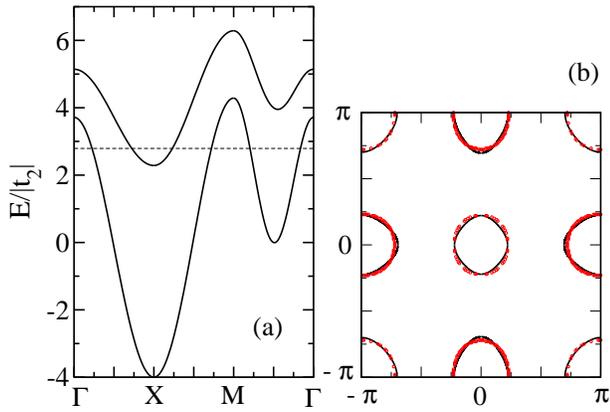}
\caption{(color online) (a) Band dispersion and (b) Fermi surface of
the half-filled two-orbital $s$-model given by
Eq.~(\ref{Sdis}) (red circles). The continuous line is the
FS for the two-orbital $d$-model.}
\vskip -0.5cm
\label{set4}
\end{center}
\end{figure}

It is clear that here both $\tau_0$ and $\tau_3$ transform like A$_{1g}$ and
for this reason we call this model the $s$-model.
In Fig.~\ref{set4}, the band
dispersion (panel (a)) and the FS (panel (b), red circles) are shown for the parameter values
$t_1=-0.05$, $t_2=0.7$, $t_3=-0.1$, $t_4=0.3$, $\Delta=2.8$ and $\mu=1.95$. The
FS of the $d$-model is also shown (continuous black line) for comparison. They are
obviously very similar, and precisely the goal of this effort is to investigate what kind
of magnetic and pairing properties emerge from these two models that have nearly
equal Fermi surfaces.

 
The hole pockets at the $\Gamma$ and $M$ points nest into the electron pockets
at $X$ and $Y$, with nesting vectors $(0,\pi)$ and $(\pi,0)$. The system is
half-filled but the individual bands/orbitals are not. Note
that this is the case with the orbitals in the multi-orbital systems proposed for
the pnictides, where nesting occurs between electron and hole pockets at the
FS but none of the orbitals is exactly half-filled.\cite{graser,kuroki}
 
\subsection{Coulomb Interaction}\label{inter}
 
The Coulomb interaction term in both Hamiltonians is the usual one, with an on-site
intraorbital (interorbital) Coulomb repulsion $U$ ($U'$), and a Hund coupling
$J$ satisfying the relation $U'$=$U-2J$ for simplicity, and a pair-hopping term with
coupling $J'$=$J$.~\cite{oles83} The full interaction term is given by
 
\begin{equation}
\begin{split}
H_{int}=
U\sum_{{\bf i},a}n_{{\bf i},a,\uparrow}n_{{\bf i},
a,\downarrow}
+{(U'-J/2)\over{2}}\sum_{{\bf i},a}n_{{\bf i},a}n_{{\bf i},-a}\\
-J\sum_{{\bf i},a}{\bf S}_{\bf{i},a}\cdot{\bf S}_{\bf{i},-a}
+{J\over{2}}\sum_{{\bf i},a}(d^{\dagger}_{{\bf i},a,\uparrow}
d^{\dagger}_{{\bf i},a,\downarrow}d_{{\bf i},-a,\downarrow}
d_{{\bf i},-a,\uparrow}+h.c.),
\end{split}
\label{int}
\end{equation}
\noindent where $d^{\dagger}_{{\bf i},a,\sigma}$ creates an electron 
with spin $\sigma$ at site ${\bf i}$ and
orbital $a=x,y$ or 1, 2. ${\bf S}_{{\bf i},a}$
($n_{{\bf i},a}$) is the spin (electronic density) of the orbital $a$ at site
${\bf i}$.
 
\section{Magnetic Properties}\label{magnetic}
 
 
For a single-orbital model, the magnetic structure factor is easily defined as
\begin{equation}
S({\bf k})=\sum_{\bf r}e^{i{\bf k.r}}\omega({\bf r}),
\label{sk}
\end{equation}
\noindent with
\begin{equation}
\omega({\bf r})={1\over{N}}\sum_{\bf i}m({\bf i})m({\bf i+r}),
\label{wr}
\end{equation}
\noindent where $N$ is the number of sites of the lattice and
\begin{equation}
m({\bf i})=n_{{\bf i},\uparrow}-n_{{\bf i},\downarrow}=
d^{\dagger}_{{\bf i},\uparrow}d_{{\bf i},\uparrow}-
d^{\dagger}_{{\bf i},\downarrow}d_{{\bf i},\downarrow},
\label{m}
\end{equation}
\noindent where $m({\bf i})$ denotes the net magnetization at site ${\bf i}$.
 
In a multiorbital system the net magnetization at site ${\bf i}$ is obtained
in terms of the magnetization of each orbital $a$, and it is given
by
\begin{equation}
m({\bf i})=\sum_a n_{{\bf i},a,\uparrow}-n_{{\bf i},a,\downarrow}=
\sum_a (d^{\dagger}_{{\bf i},a,\uparrow}d_{{\bf i},a,\uparrow}-
d^{\dagger}_{{\bf i},a,\downarrow}d_{{\bf i},a,\downarrow}).
\label{mmulti}
\end{equation}
 
While Eq.~\ref{mmulti} characterizes the magnetization that is measured in experiments such as neutron scattering,
it is natural to define generalized magnetic moments
$m_{ab}({\bf i})$\cite{dhlee} given by
\begin{equation}
m_{ab}({\bf i})=d^{\dagger}_{{\bf i},a,\uparrow}d_{{\bf i},b,\uparrow}-
d^{\dagger}_{{\bf i},a,\downarrow}d_{{\bf i},b,\downarrow}.
\label{mab}
\end{equation}
 
\begin{figure}[thbp]
\begin{center}
\includegraphics[width=7.6cm,clip,angle=0]{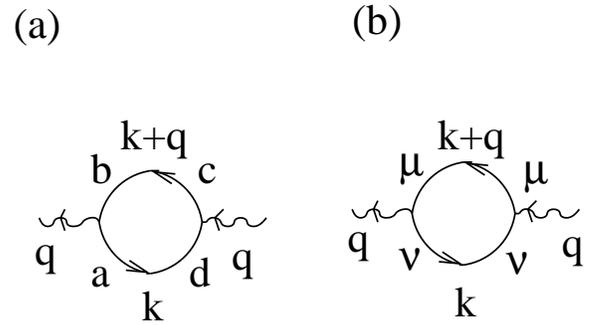}
\caption{(a) Electronic process that gives rise
to the orbital components of the structure factor. (b) Same as (a) but in the
band representation.}
\vskip -0.5cm
\label{bubble}
\end{center}
\end{figure}

With this definition, a generalized form
of the magnetic correlation functions will depend on 4 orbital indices:
\begin{equation}
\omega_{abcd}({\bf r})={1\over{N}}\sum_{\bf i}m_{ab}({\bf i})m_{cd}({\bf i+r}).
\label{wrabcd}
\end{equation}
Thus, it is possible to define  orbital dependent magnetic structure factors
given by:
\begin{equation}
S_{abcd}({\bf k})=\sum_{\bf r}e^{i{\bf k.r}}\omega_{abcd}({\bf r}).
\label{skabcd}
\end{equation}
These orbital-dependent operators may arise from processes as those depicted
in panel (a) of Fig.~\ref{bubble}, where having different orbitals at the
two vertices is possible if the orbitals strongly hybridize to form a
band.\cite{kemper}
 
The total orbital magnetic structure factor can then be defined as:
\begin{equation}
S_{TO}({\bf k})=\sum_{a,b,c,d}S_{abcd}({\bf k}).
\label{skt}
\end{equation}
                                                                                
Note that there are $M^4$ orbital dependent components of the
generalized magnetic structure factor, where $M$ is the number of active 
orbitals in the  
system. The magnetization that is measured in neutron scattering experiments
is given by Eq.~\ref{mmulti}, which in terms of the components of the
tensor $m_{ab}$ becomes
\begin{equation}
m({\bf i})=\sum_a n_{{\bf i},a,\uparrow}-n_{{\bf i},a,\downarrow}=
\sum_a m_{aa}({\bf i})=tr [m_{ab}({\bf i})].
\label{mmultia}
\end{equation}
Since $m({\bf i})$ is a trace its value is independent of the basis chosen to define the orbitals and
it allows to calculate the experimentally measured local magnetization. 

Notice that $m({\bf i})$ is the operator 
that has to be considered  in order to construct the so-called homogeneous or diagonal structure
factor defined in terms of the diagonal (intra-orbital) magnetic moments
$m_{aa}({\bf i})$ and given by \cite{graser,moreo}
\begin{equation}
S_{MO}({\bf k})={1\over{N}}\sum_{a,b,{\bf r},{\bf i}}
e^{i{\bf k.r}}m_{aa}({\bf i})m_{bb}({\bf i+r})
=\sum_{a,b}S_{aabb}({\bf k}).
\label{skMO}
\end{equation}
$S_{MO}$ is the physical magnetic structure factor that has to be calculated in the
context of multiorbital systems to compare with neutron scattering results.\cite{Daghofer:2009p1970,moreo} 
Several authors have pointed out the existence of the generalized components of
the magnetic susceptibility both in the orbital representation\cite{graser,maria} and in
the band representation.\cite{phil} It has also been pointed out that an orbital-transverse density-wave (OTDW)
ordered state characterized by the non-homogeneous components of the magnetization tensor may
develop in multi-orbital systems,\cite{yao} an issue that will be further explored and discussed 
in the present work.

\subsection{Non-interacting case}

In order to understand the
relationship between $S_{TO}$, $S_{MO}$, and the properties of the
FS of the system, it is illuminating to consider the
non-interacting case which can be easily studied in momentum space. Via a Fourier transform of
$d^{\dagger}_{{\bf i},a,\sigma}$ and $d_{{\bf i},a,\sigma}$, $S_{abcd}$ in Eq.~\ref{skabcd} can be written as
\begin{equation}
S_{abcd}({\bf k})=\sum_{{\bf p},{\bf q},\sigma,\sigma'}(-1)^{\sigma+\sigma'}d^{\dagger}_{{\bf q},a,\sigma}d_{{\bf q+k},b,\sigma}
d^{\dagger}_{{\bf p},c,\sigma'}d_{{\bf p-k},d,\sigma'}.
\label{skabcdk}
\end{equation}
In momentum space it is natural to use the band representation in which
\begin{equation}
\begin{split}
S_{abcd}({\bf k})=\sum_{{\bf p},{\bf q},\sigma,\sigma',\mu,\mu',\nu,\nu'}(-1)^{\sigma+\sigma'}\\
<\mu|a>_{\bf q}<b|\mu'>_{\bf q+k}<\nu|c>_{\bf p}<d|\nu'>_{\bf p-k}\\
d^{\dagger}_{{\bf q},\mu,\sigma}d_{{\bf q+k},\mu',\sigma}
d^{\dagger}_{{\bf p},\nu,\sigma'}d_{{\bf p-k},\nu',\sigma'},
\end{split}
\label{skabcdkb}
\end{equation}
\noindent where $d^{\dagger}_{{\bf p},\nu,\sigma}$ creates an electron with momentum ${\bf p}$ and 
$z$-spin component
$\sigma$ at band $\nu$,  while $<\nu|a>_{\bf p}$ is the matrix element 
for the transformation from orbital to band representation.
 
In the band representation, 
the electronic processes that contribute to the magnetic correlations are shown in
panel (b) of Fig.~\ref{bubble}. Since the electronic band cannot change as the 
electron created at the right vertex is
destroyed at the left vertex, in the band representation 
we can define band-dependent components of the structure
factor given by
\begin{equation}
S_{\mu\nu\nu\mu}({\bf k})=\sum_{{\bf p},{\bf q},\sigma} d^{\dagger}_{{\bf q},\mu,\sigma}d_{{\bf q+k},\nu,\sigma}
d^{\dagger}_{{\bf p},\nu,\sigma}d_{{\bf p-k},\mu,\sigma},
\label{skmunub}
\end{equation}
\noindent where the greek indices label the bands. 
A total structure factor can be defined in terms of $S_{\mu\nu\nu\mu}$ as
\begin{equation}
S_{TB}({\bf k})=\sum_{\mu,\nu}S_{\mu\nu\nu\mu}({\bf k}).
\label{sktb}
\end{equation}
\noindent Also the homogeneous or diagonal magnetic structure 
factor $S_{MB}$, analogous of $S_{MO}$, can be defined as
\begin{equation}
S_{MB}({\bf k})=\sum_{\mu}S_{\mu\mu\mu\mu}({\bf k}),
\label{skMB}
\end{equation}
\noindent since in the band representation $S_{\mu\mu\nu\nu}=0$, if $\mu\ne\nu$.
Note that the band representation is the natural starting point in approaches
based on fermiology.\cite{mazin,chubu}
 
In the noninteracting case being considered in this section, it is easy to show that
\begin{equation}
S_{\mu\nu\nu\mu}({\bf k})=2\sum_{\bf q} f_{\mu}({\bf q})[1-f_{\nu}({\bf q+k})],
\label{skmunubf}
\end{equation}
\noindent where $f_{\mu}({\bf q})$ is the Fermi function for the band $\mu$. We also find that
the components of the structure factor in the orbital representation are given by
\begin{equation}
\begin{split}
S_{abcd}({\bf k})=
2\sum_{{\bf q},\mu,\nu}
<\mu|a>_{\bf q}<b|\nu>_{\bf q+k}\\<\nu|c>_{\bf q+k}<d|\mu>_{\bf q}
f_{\mu}({\bf q})[1-f_{\nu}({\bf q+k})].
\end{split}
\label{skabcdkbf}
\end{equation}
From the expressions in Eqs.~\ref{skmunubf} and \ref{skabcdkbf} it can be shown 
that $S_{TO}=S_{TB}$ and $S_{MO}=S_{M B}$ only if the orbitals do not hybridize 
to form the bands, i.e., the matrix elements are the elements of the identity matrix. 
In case of a nonzero hybridization, then
the structure factors in the band and orbital representations are different.

\subsection{$d$-model}\label{do}
                                                                                
\begin{figure}[thbp]
\begin{center}
\includegraphics[width=8.0cm,clip,angle=0]{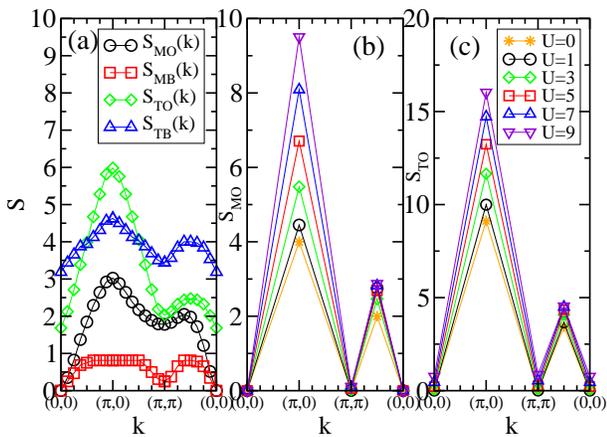}
\caption{(color online) (a) Magnetic structure factors, total and homogeneous as
indicated, for
the non-interacting two-orbital $d$-model on a $16 \times 16$
lattice. 
(b) Homogeneous orbital magnetic structure factor $S_{MO}({\bf k})$ for the interacting case
with $J/U=0.25$ and at the indicated values of $U$. The results were obtained
numerically using an 8-sites cluster and the Lanczos method. (c) Total 
orbital magnetic structure factor $S_{TO}({\bf k})$ for the interacting case, for
the same parameters and technique as used in (b).}
\vskip -0.5cm
\label{dsk}
\end{center}
\end{figure}

Numerical Lanczos calculations for the homogeneous (or diagonal) magnetic structure factor $S_{MO}$
have already been performed in previous literature 
for the two-orbital $d$-model indicating a tendency
towards a magnetic stripe ordering for the undoped
case, characterized by peaks at ${\bf k}=(\pi,0)$ and $(0,\pi)$ in $S_{MO}$.\cite{moreo}
This tendency is already apparent even in the non-interacting case\cite{Daghofer:2009p1970,moreo}
as illustrated in panel (a) of Fig.~\ref{dsk} where $S_{MO}$ calculated in a $16\times 16$
cluster is shown with open circles, along the directions $(0,0)-(\pi,0)-(\pi,\pi)-(0,0)$ in
the unfolded BZ. The broad peak at ${\bf k}=(\pi,0)$ is clear and it can be compared with the curve
denoted by the star symbols in panel (b) of the same figure where results for the  
$\sqrt{8}\times\sqrt{8}$ cluster that can 
be studied numerically exactly (with the Lanczos algorithm and for any value of the Hubbard
couplings) are presented. This same behavior is
also apparent in the total orbital structure factor $S_{TO}({\bf k})$ indicated
by the diamonds
in Fig.~\ref{dsk}(a). 

On the other hand, a calculation of the magnetic structure factor 
using the band representation, i.e.
$S_{MB}({\bf k})$ indicated by the squares in panel (a) of Fig.~\ref{dsk}, shows a rather different
behavior:
instead of a clear peak at $(\pi,0)$ there is a featureless plateau 
around $(\pi,0)$ that extends to
$(\pi/2,\pi/2)$. This example demonstrates the importance 
of the matrix elements in Eq.~\ref{skabcdkb} which
differentiate between $S_{MO}$ and $S_{MB}$. In the non-interacting case, 
both functions can be expressed
in terms of the Fermi functions as in Eqs.~\ref{skmunubf} and \ref{skabcdkbf} allowing us to conclude
that the peak at $(\pi,0)$ arises from the matrix elements rather than from purely nesting effects
of the Fermi surfaces. Ignoring the matrix elements, it is interesting 
to note that a feature at $(\pi,0)$ can also develop 
if all the components of the structure factor in the band representation are
considered and $S_{TB}({\bf k})$ is calculated, as shown by the curve indicated with triangles in
Fig.~\ref{dsk}(a).

\begin{figure}[thbp]
\begin{center}
\includegraphics[width=8.0cm,clip,angle=0]{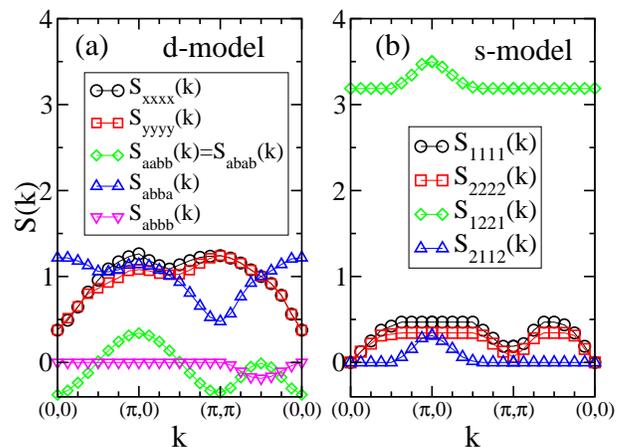}
\caption{(color online) (a) Orbital-resolved components of the magnetic structure factor for the
non-interacting two-orbital $d$-model on a $16 \times 16$ lattice. (b) Band/orbital resolved components of the 
magnetic structure factor for the non-interacting two-orbital $s$-model also on a
$16 \times 16$ lattice. The index 1 (2) labels the lower (upper) band.}
\vskip -0.5cm
\label{skcomp}
\end{center}
\end{figure}

The contribution of the band- and orbital-resolved components of the structure factor in the
non-interacting case are presented in panel (a) of Fig.~\ref{skcomp}.
The components of the structure factor that contribute to $S_{MO}$ are $S_{aabb}$ with $a$ ($b$) taking the values
$x$ ($y$) and $y$ ($x$) indicated by the diamonds in the figure, and $S_{aaaa}$ (indicated by the circles and squares).
It is clear from the figure that the peak at $(\pi,0)$ in $S_{MO}$ at the non-interacting level is mostly due to
the $S_{aabb}$ that arise from the nesting of the two bands that contain the same orbital flavors due to hybridization, while
the components of the form $S_{aaaa}$ show features also at $(\pi,\pi)$ since this wave vector also nests the hole (electron) FS's at
$\Gamma$ and $M$ ($X$ and $Y$).
It can be seen that the non-homogeneous components of the form $S_{abab}$ (diamonds) behave as $S_{aabb}$ in the non-interacting case 
and contribute to form the peak at $(\pi,0)$ in the total structure factor $S_{TO}$ [triangles in Fig.~\ref{dsk}(a)].
For completeness in Fig.~\ref{skcomp}(a) orbital resolved structure factors of the form $S_{abba}$ (up triangles) and 
$S_{abbb}$ (down triangles) are also shown; $S_{abba}$ increase the value of $S_{TO}$ at $(\pi,0)$ while
$S_{abbb}$ provide a small negative
contribution to $S_{TO}$ along the diagonal direction of the BZ. Similar 
results were obtained for all the correlations in which three of the four indices
are the same.

In non-interacting single-orbital systems, as studied for the cuprates, 
the spin and charge susceptibilities have the same form
for all values of non-zero momenta, and any features in these functions 
arise from the nesting properties of the
Fermi surface. Naively, the same is expected in the case of multi-orbital 
models but, as it will be discussed below, the
hybridization of the orbitals plays a crucial role.
In the $d$-model, the peaks in $S_{MO}$ appear to be associated 
with the nesting of the hole- and electron-like Fermi surfaces.
In the weak coupling picture, it is expected that magnetic order with
${\bf Q}$ equal to the nesting moments stabilizes when
repulsive Coulomb interactions are added. Our
Lanczos calculations for $S_{MO}$ and $S_{TO}$, in panels (b) and (c) of Fig.~\ref{dsk}, show that this is indeed the case.
 
The Lanczos calculated orbital magnetic
structure factor $S_{MO}({\bf k})$, using a $\sqrt{8}\times\sqrt{8}$ sites cluster, 
is shown in Fig.~\ref{dsk}(b) for different values of $U$ and
at $J/U=0.25$. This structure factor has a peak
at ${\bf k}=(\pi,0)$ (and $(0,\pi)$ as well, not shown)
that becomes sharper as $U$
increases, indicating a tendency towards robust 
magnetic order. Mean-field calculations
based on these results, but
extended to much larger systems, indicate that actual magnetic order 
develops at a finite value of $U$.\cite{moreo,rong}
 
\begin{figure}[thbp]
\begin{center}
\includegraphics[width=8.0cm,clip,angle=0]{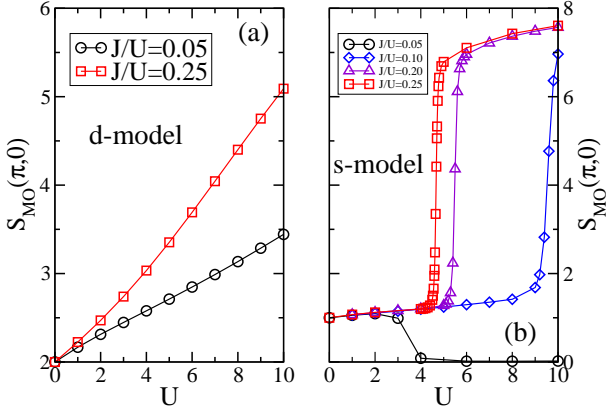}
\caption{(color online) Orbital magnetic structure factor at wave vector $(\pi,0)$
calculated numerically (Lanczos). (a) Results for the two-orbital $d$-model, as a
function of the Coulomb repulsion $U$ and for the values of $J/U$ indicated.
(b) Same as (a) but for the $s$-model.}
\vskip -0.5cm
\label{skd}
\end{center}
\end{figure}

The Lanczos-evaluated behavior of the 
$S_{MO}({\bf k})$ peak at ${\bf k}=(\pi,0)$, as a function of $U$, is shown in
Fig.~\ref{skd}(a), for two different values of $J/U$ (0.05 and 0.25). The tendency towards a robust 
magnetic state with increasing $U$ and $J/U$ is again clear.
 
\begin{figure}[thbp]
\begin{center}
\includegraphics[width=8.0cm,clip,angle=0]{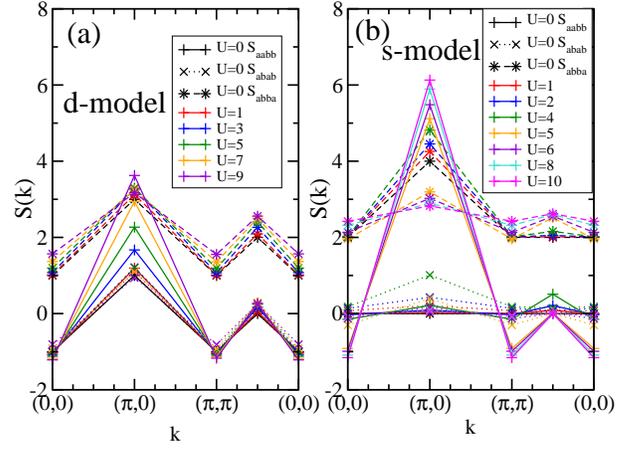}
\caption{(color online) Orbital-resolved components of the total structure factor (sums over repeated indices are implied): $S_{aabb}$ (plus, full line),
$S_{abab}$ (x, dotted line) and $S_{abba}$ (star, dashed line), for the values of $U$ indicated,
obtained numerically (Lanczos) 
at $J/U=0.25$ using an 8-sites cluster for (a) the $d$-model and (b) the $s$-model.}
\vskip -0.5cm
\label{resolved}
\end{center}
\end{figure}

As previously stated, $S_{MO}$ is the magnetic structure factor calculated in 
the literature for comparison with experiments, but for completeness and for 
the sake of comparison with the $s$-model results, in panel (c) of Fig.~\ref{dsk},
we present the Lanczos calculated values for the total generalized magnetic
moment $S_{TO}$ for the $d$-model as a function of $U$, for the case $J/U=0.25$. It is clear that
for the $d$-model $S_{TO}$ mimics the behavior of $S_{MO}$. An important question to ask
is what are the components of the orbital-resolved magnetic structure factor that
drive the development of a 
peak at ${\bf Q}=(\pi,0)$ (and $(0,\pi)$) when the Coulomb interactions are active. In Fig.~\ref{resolved} partial sums over selected components
of the structure factor are shown with summations performed over repeated indices. In panel (a) of Fig.~\ref{resolved}
it can be clearly observed that $S_{aabb}$, whose sum over $a$ and $b$ are indicated by the plus signs and the continuous lines in different shades for the different 
values of the interaction, 
are the components that drive that magnetic behavior.
In fact, these are the homogeneous components that contribute to the physical
magnetic structure factor $S_{MO}$. It is interesting to note that 
while $\sum_{a,b}S_{aabb}$ is equal to $\sum_{a,b}S_{abab}$
in the non-interacting system (panel (a) of Fig.~\ref{skcomp}) the
partial sum of the non-homogeneous component $\sum_{a,b}S_{abab}$ [x symbols and dotted lines in 
Fig.~\ref{resolved}(a)] does not increase with $U$ at ${\bf Q}$ while the partial sum $\sum_{a,b}S_{aabb}$ clearly does.

\subsection{$s$-model}\label{so}
 
Let us now carry out a similar analysis but for the two-orbital $s$-model defined
by Eq.~\ref{Sdis}. Since in this model each band is defined by a single orbital, then
it is clear that $S_{MO}=S_{MB}$ and $S_{TO}=S_{TB}$.\cite{foot}
Note that studies based on fermiology assume that if hole and electron FS's are nested via
a momentum vector ${\bf Q}$, then spin density wave order 
will arise from a logarithmic instability
that develops in the spin response at ${\bf Q}$ and is stabilized by the Coulomb interaction.\cite{mazin,chubu} 
In this scenario the
spin-density wave originates from the formation of particle-hole pairs, excitons,
belonging to the electron and hole FS's (excitonic mechanism).\cite{chubu} 
Our goal is to investigate whether this mechanism is valid for the $s$-model.
 
\begin{figure}[thbp]
\begin{center}
\includegraphics[width=8.0cm,clip,angle=0]{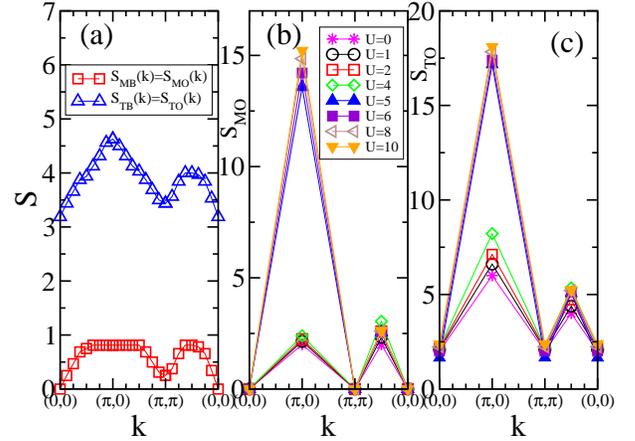}
\caption{(color online) (a) Magnetic structure factors (total and homogeneous)
as indicated for the non-interacting two-orbital $s$-model using a
$16 \times 16$ lattice. (b) Homogeneous orbital/band magnetic structure
factor $S_{MO}({\bf k})$ for the interacting case
with $J/U=0.25$, at the indicated values of $U$. The results were obtained
numerically via the Lanczos method using an 8-sites cluster. (c) Total 
orbital/band magnetic structure
factor $S_{TO}({\bf k})$ for the interacting case with the same parameters as in (b).
}
\vskip -0.5cm
\label{s0sk}
\end{center}
\end{figure}
 
The magnetic structure factor $S_{MO}$ in the non-interacting limit,
denoted by the squares in panel (a) of
Fig.~\ref{s0sk}, does not show the features expected from the nesting of the
two Fermi surfaces at momentum ${\bf Q}$. The structure factor 
 is actually rather flat on all the BZ,
vanishing at ${\bf k}=(0,0)$ and 
$(\pi,\pi)$. These results are 
not what it would have been expected from the nesting properties.

Note that the results for $S_{MO}$ in the non-interacting $s$-model 
[squares in Fig.~\ref{s0sk}(a)] are actually identical to the results for the homogeneous structure factor in the
$d$-model in the band representation $S_{MB}$ [indicated by squares in Fig.~\ref{dsk}(a)], since both systems 
do have the same FS. However, note how
different are the results for the $d$-model in the orbital 
representation [indicated by circles in Fig.~\ref{dsk}(a)]. This is due to the effect
of the matrix elements that result from the hybridization 
of the orbitals, which play a crucial
role in the magnetic properties of the system. This effect 
can be more clearly appreciated when the interactions are added.
The behavior of the peak in $S_{MO}({\bf k})$ at
${\bf k}=(\pi,0)$ was calculated with the Lanczos method applied to the $s$-model, 
by varying $U$ and at different values of $J/U$ using an $N=8$ sites tilted cluster. In
Fig.~\ref{skd}(b) it can be observed that for values
of $J<0.1U$ the peak in $S_{MO}$ eventually vanishes. On the other hand,
for $J\ge 0.1U$ a
rapid increase in the peak's magnitude suddenly occurs at a value of $U$ that decreases as
$J/U$ increases. The increase of the peak at $(\pi,0)$ with increasing $U$ is contrasted
with the behavior of the feature at $(\pi/2,\pi/2)$ displayed in Fig.~\ref{s0sk}(b).
Examination of the numerical (Lanczos) ground state indicates that at
this point the Hubbard interaction is strong enough to hybridize the two bands
and develop magnetic stripe order. 

\begin{figure}[thbp]
\begin{center}
\includegraphics[width=8.0cm,clip,angle=0]{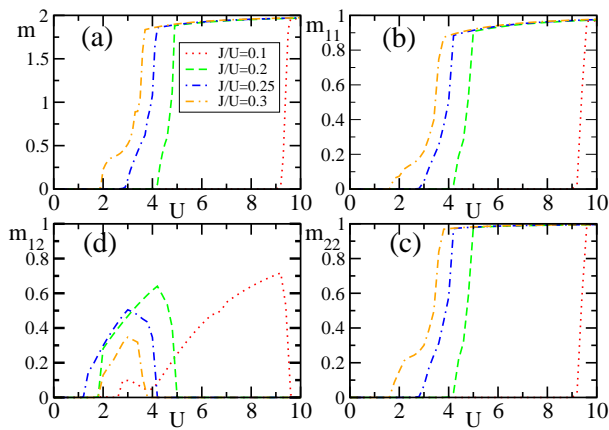}
\caption{(color online) Mean-field calculated
orbital/band  resolved magnetic order parameters for the $s$-model, as a function of $U$
and for the indicated values of $J/U$. (a) Total homogeneous magnetic order parameter $m=m_{11}+m_{22}$; 
(b) $m_{11}$; (c) $m_{22}$; (d) $m_{12}=m_{21}$.} 
\vskip -0.5cm
\label{mfs}
\end{center}
\end{figure}

\begin{figure}[thbp]
\begin{center}
\includegraphics[width=8.0cm,clip,angle=0]{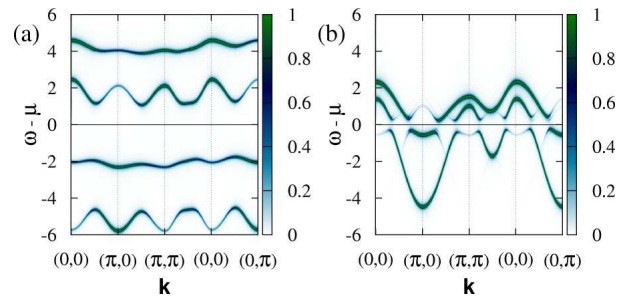}
\caption{(color online) Intensity of the mean-field calculated spectral functions $A({\bf k},\omega)$ as a function of  
$\omega-\mu$ and ${\bf k}$ for the $s$-model:
(a) in the stripe magnetic phase for $U=5$ and $J/U=0.25$; (b) in the phase with orbital-transverse spin order
for $U=2.5$ and $J/U$=0.25.}
\vskip -0.5cm
\label{akw}
\end{center}
\end{figure}

Based on the numerical results discussed above a Hartree-Fock mean-field calculation was 
performed, 
following technical aspects already widely discussed
in previous literature.\cite{rong} By this procedure we
found that the total (homogeneous) magnetization $m$ shown in panel (a) of Fig.~\ref{mfs}
mimics the behavior of $S_{MO}(\pi,0)$. Here, the transition to the magnetically ordered
state is very rapid, resembling a first-order transition. We observed that MF magnetic order develops only if 
$J\ge 0.1U$ which is in agreement with the Lanczos results shown in Fig.~\ref{skd}(b).   
The mean-field results also indicate that a full gap
characterizes the magnetic state which is then an insulator as it can be seen from the MF calculated 
spectral functions $A({\bf k},\omega)$ displayed on panel (a) of Fig.~\ref{akw}. It is clear that the
hybridization of the original bands/orbitals due to the Coulomb interaction is very strong and the band
structure has been totally reconstructed.
This behavior can be understood in the real-space representation. In
order to develop magnetic stripes in the half-filled system, 
it is necessary to have a net magnetic moment on each site. In the
$d$-model, each orbital is half filled and thus contains a spin-$1/2$
that can easily be polarized by the interaction. In the $s$-model, on
the other hand, the orbitals correspond to the bands, and one orbital
is thus almost filled while the other is almost empty. Then, there are
far fewer magnetic moments that can be polarized.

Thus, we observe that in the $s$-model the peak at ${\bf Q}$ in the
magnetic structure factor does not develop from the nesting of the FS but from the
Coulomb interaction, and it occurs fairly suddenly and at a robust value of $U\ge 4$ for 
the hopping parameters used here.
Thus, while nesting appears to be a needed condition for the development of
the peak in the magnetic structure factor, it is not a sufficient condition.
The hybridization of the orbitals needs to be present such
that the matrix elements allow the peak to emerge
at sufficiently strong coupling.
In fact, it is necessary that the bands that are connected 
by the nesting vector ${\bf Q}$ share
the same orbital flavor. If this occurs 
via hybridization, magnetic order can develop at relatively weak
coupling, but if this is not the case, the Coulomb interaction 
would induce magnetic order
only in the strong coupling regime, as we have 
verified by studying the $s$-model. 
In this case, the magnetic transition is also a metal-insulator transition, as observed 
at least within the mean-field approximation.
The $d$-orbital model, on the other hand, is known to display an intermediate metallic
magnetic phase.\cite{rong} Thus, the present results indicate that the $s$ and
$d$ models develop similar magnetic behavior only in the strong
coupling regime while in weak coupling, 
despite the nearly identical Fermi surfaces, both models have
quite different ground states.

\subsubsection{Orbital-transverse spin order}

While the analysis of the results for the $s$-model presented above indicates that, despite the nesting of the
electron and hole FS's, no magnetic order, as defined by the homogeneous operator, develops in weak coupling, 
it is instructive to analyze the behavior of the non-homogeneous components and the total magnetic structure factor
$S_{TO}$. The non-interacting values of $S_{TO}$ on a $16 \times 16$ lattice are indicated by the triangles
in panel (a) of Fig.~\ref{s0sk}. 
There is a feature at $(\pi,0)$ arising from 
the contribution of the interband components 
of the form $S_{abba}\equiv S_{\mu\nu\nu\mu}$, shown by the triangles and
diamonds in panel (b) of Fig.~\ref{skcomp}. These are the components of $S_{abcd}$ that contribute
to the development of the maximum at ${\bf Q}=(\pi,0)$ (and $(0,\pi)$) because the nesting at ${\bf Q}$
is between FS's defined by different bands. However, this type of terms are not part
of the definition of the homogeneous structure factor $S_{MB}$.
On the other hand, the components of the form $S_{aaaa}$ indicated with circles and squares in Fig.~\ref{skcomp}(b), 
have a very flat shape
in all the BZ and do not produce a sharp feature. Any other
combination of orbital indices does not contribute to $S_{TO}$ as shown in Eq.~\ref{skmunubf}.

\begin{figure}[thbp]
\begin{center}
\includegraphics[width=8.0cm,clip,angle=0]{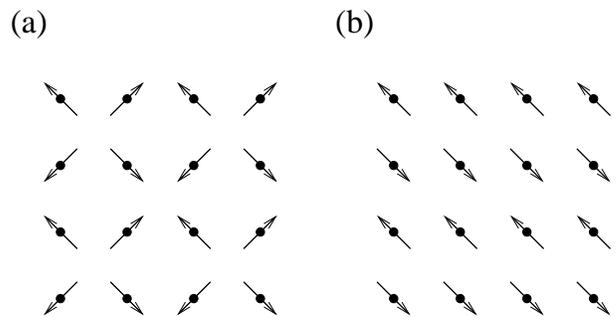}
\caption{Schematic representation of the real space mean-field calculated
ground states for the $s$-model when $m_{12}$ is non-zero. (a) Flux phase; (b) Stripe phase.
The dots indicate the sites and the arrows represent the MF value of the 
generalized spin ${\bf G}_{12}({\bf i})$ defined in the text.} 
\vskip -0.5cm
\label{cartoon}
\end{center}
\end{figure}

The effect of the Coulomb interactions on the feature at $(\pi,0)$ in $S_{TO}$ has been obtained
with Lanczos calculations and it can be seen in panel (c) of Fig.~\ref{s0sk}. The peak slowly 
increases as $U$ raises from 0 to 4. Notice that for the same range of values of $U$ the peak in
$S_{MO}$ shown in panel (b) of the figure does not change. The obvious question is whether this behavior indicates
a novel kind of order in multi-orbital systems. The answer is provided via our MF approach that allows us to evaluate
the components of the magnetization $m_{ab}$. The homogeneous magnetization $m$ displayed in panel (a) of Fig.~\ref{mfs}
is obtained as the sum of the intraorbital magnetizations $m_{11}$ and $m_{22}$ shown in panels (b) and (c) of the
figure. Interestingly, we found that the non-diagonal components $m_{12}=m_{21}$ develop finite 
values while the diagonal components are zero
for values of $J/U>0.1$ as shown in panel (d) of the figure. At the MF level we can study the real space configuration associated to
this finite order parameter. We have observed that the orbital spins are 
disordered, which is expected by the lack of features in $S_{MO}({\bf k})$, but there are ordered generalized spins ${\bf G}_{ab}({\bf i})$ defined
as
\begin{equation}
{\bf G}_{ab}({\bf i})=d^{\dagger}_{{\bf i},a,\alpha}\vec\sigma_{\alpha,\beta}d_{{\bf i},b,\beta},
\label{G}
\end{equation}
\noindent where $\vec\sigma$ are the Pauli matrices and the orbital indices $a\ne b$. 
In Fig.~\ref{cartoon} we show two configurations of ${\bf G}_{12}({\bf i})$ that provide the MF ground state associated
with the peak in $S_{TO}$ at $(\pi,0)$ (and $(0,\pi)$) when $m_{12}$ is finite. Panel (a) shows a flux configuration that generates
peaks at $(\pi,0)$ and $(0,\pi)$ in $S_{TO}$ and panel (b) shows a stripe configuration that produces a peak at $(0,\pi)$. 
The peak at $(\pi,0)$ is generated by a companion configuration rotated by $\pi/2$.
Flux and stripe configurations have energies very close to each other and the actual ground state depends on the parameters.\cite{eremin}  

The new phase hinted at by the Lanczos calculations and stabilized in the MF calculations 
is insulating. The MF calculated spectral functions are shown in panel
(b) of Fig.~\ref{akw}. A full gap has developed at the FS indicating that this order, if realized, would be observed with ARPES measurements. On the other hand,
neutron scattering experiments would not detect it. This can be seen by performing a rotation in orbital space given by\cite{yao}
\begin{equation}
d^{\dagger}_{{\bf i},\pm,\sigma}={1\over{\sqrt{2}}}(d^{\dagger}_{{\bf i},1,\sigma}\pm d^{\dagger}_{{\bf i},2,\sigma}).
\label{pm}
\end{equation}

\begin{figure}[thbp]
\begin{center}
\includegraphics[width=8.0cm,clip,angle=0]{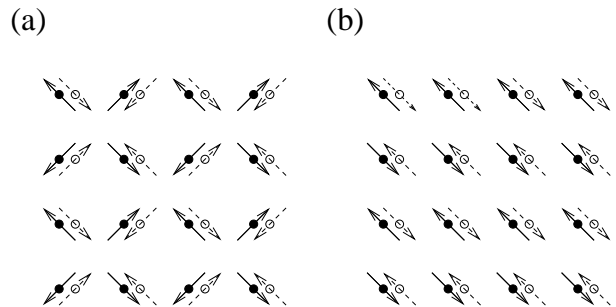}
\caption{Schematic representation of the real space mean-field calculated
ground states for the $s$-model when $m_{12}$ is non-zero: (a) Flux phase; (b) Magnetic stripe phase.
The black and white dots represent the orbitals + and - at each site and the continuous 
and dashed arrows represent the MF value of the 
spin at each orbital.} 
\vskip -0.5cm
\label{cartoon2}
\end{center}
\end{figure}

In this new basis the schematic representations of the spins are shown in Fig.~\ref{cartoon2}. It is clear that while the homogeneous spins in the orbitals + (black dots)
and - (white dots) are ordered, the net spin at each site is 0 and thus, neutron scattering experiments will not detect the order because there is no finite local magnetization. These phases appear to be a realization of the
orbital-transverse density-wave (OTDW) order proposed in Ref.~\onlinecite{yao}.

Summarizing, 
a careful analysis of the small-cluster ground states obtained via Lanczos techniques, 
and with mean-field approximations in larger clusters, highlights the important 
role that the orbital composition
plays in the development of magnetic order.

For the $s$-model, it is illuminating to consider the behavior of the total magnetic structure factor
$S_{TO}$, see panel (c) of Fig.~\ref{s0sk}, calculated numerically as the interactions are increased. 
There is a weak increase of $S_{TO}$ at ${\bf Q}$ before the sudden jump at $U=4$. The behavior
of the resolved components displayed in panel (b) of Fig.~\ref{resolved} shows that for $0\le U\le 4$
the partial sum over $a$ and $b$ of the non-homogeneous components $S_{abab}$ (x symbols and dotted line) and $S_{abba}$ (star symbols and dashed lines) 
increases in value at $(\pi,0)$ indicating the stabilization
of the orbital-transverse spin phase.  For $U>4$ a sudden increase
of the sum of the homogeneous components 
$S_{aabb}$ (plus symbols and continuous line) develops, the non-homogeneous components start
to decrease and homogeneous magnetic order is established.

\subsection{Weak Coupling: RPA Analysis}
 
\begin{figure}[thbp]
\begin{center}
\includegraphics[width=8.0cm,clip,angle=0]{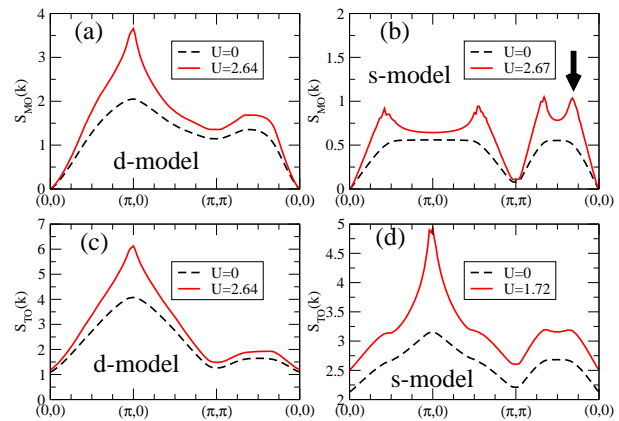}
\caption{(color online) RPA calculated magnetic structure factors for
$J/U=0.25$, at the values of $U$ indicated (full line). The non-interacting
results are indicated with dashed lines. (a) Homogeneous magnetic structure
factor in the $d$-model. (b) Same as (a), but for the $s$-model. 
The arrow indicates the peak that grows the most as the
critical $U$ is reached. (c) Generalized magnetic structure factor for the $d$-model. 
(d) Same as (c), but for the $s$-model.}
\vskip -0.5cm
\label{rpa}
\end{center}
\end{figure}

Additional insight into the weak coupling behavior of the $d$- and $s$-models can
be obtained via the diagrammatic RPA method. Using this technique, the
magnetic susceptibility $\chi_{abcd}({\bf k},i \omega)$ was
calculated,~\cite{graser,kemper} and the static structure factor was obtained
by integrating the results over $\omega$.\cite{ogata} 
In panel (a) of Fig.~\ref{rpa}, 
the RPA-calculated diagonal or homogeneous structure factor for the $d$-model 
is presented. The non-interacting result (in agreement with the results 
indicated by the circles in panel (a) of Fig.~\ref{dsk}) are denoted 
by the dashed line, while
results at $U=2.64$, the coupling strength where divergent behavior is about
to occur for the case $J/U$=$0.25$, are indicated by the continuous line.
In these results the peak at $(\pi,0)$ is very prominent both with and without
the Hubbard interaction on.

The same calculation performed for the
$s$-model, presented in panel (b) of Fig.~\ref{rpa}, gives rather different
results. The flat behavior in the noninteracting case (dashed line), in
agreement with the curve indicated by squares 
in Fig.~\ref{s0sk}(a), is replaced within RPA by a curve (continuous line) that develops
weak features at incommensurate values of the momentum. Note that there were no precursors
of these features in the non-interacting limit. 
Eventually the peak the
closest to the $\Gamma$ point along the diagonal direction of the BZ,
indicated with an arrow in the figure, was found to diverge when $U$ becomes larger than 2.67
for $J/U=0.25$. 
This appears to be an illustration of a case in which RPA calculations
indicate magnetic behavior that is unrelated to nesting  properties.
The RPA results show that an excitonic weak-coupling picture in which
magnetic order characterized by the nesting momentum ${\bf Q}$ is expected to
occur can be misleading if the orbital composition of the bands is not incorporated
into the discussion. In the excitonic picture, the expectation is that the
Coulomb interaction will allow the formation of electron-hole pairs with the
electron (hole) in the electron (hole) Fermi surface. Since $S_{MO}$ incorporates
intraorbital electron-hole pairs, an RPA response requires that the nesting
vector connects parts of the electron and hole bands that contain the {\it same}
orbital flavor. This is the case in the $d$-model where
even in the weak coupling regime the $(\pi,0)$ magnetic-stripe 
state with two electrons
with parallel spins at every site of the lattice has the largest weight in
the ground state according to our Lanczos numerical studies. Since both orbitals are
degenerate, the energetic penalization for populating both orbitals is $U'$
and there is a gain given by $J$ if both spins are parallel.
As discussed before, in the $s$-model, on the other hand,
the orbitals are non-degenerate and, thus, in addition to $U'$ 
there is an energy $\Delta$ of penalization 
when two electrons are located in different
orbitals at the same site. This energy 
can be larger than the gain obtained from $J$ by having
parallel spins or than the $U$ penalization that arises from introducing both
electrons in the same orbital. Then, a magnetic ``stripe'' state can only develop
when $U$ is comparable to the splitting $\Delta$. 
This regime, which develops in strong coupling according to our Lanczos and MF calculations, 
is not captured by the weak-coupling RPA method.
However, it will be shown that RPA is effective at finding the orbital-transverse spin state
presented in the previous section if the generalized structure factor $S_{TO}$ is calculated.

The values of $S_{TO}$ obtained with RPA are presented in
panels (c) and (d) of Fig.~\ref{rpa} for the $d$ and $s$ models, respectively. 
Both develop a peak at the nesting wavevector. 
The generalized structure factor takes into
account electron-hole pairs formed by an electron and a hole in different
bands that are allowed to have different orbital flavors. This is the reason
why a peak develops now in both cases. While in the case of the $d$-model
the behavior of $S_{TO}$ mimics $S_{MO}$ and the divergence in both occurs at the same value of $U$
(slightly above 2.64 for $J/U=0.25$) indicating that the
stripe-magnetic order is the cause, in the $s$-model the peak in $S_{TO}$ develops at a lower value of
$U$ ($U=1.72$ for $J/U=0.25$) and it results from the ordering revealed by the inhomogeneous components
$S_{1221}$ and $S_{2112}$ of the structure factor, i.e., orbital-transverse
spin order as discussed in the previous section. In this new light, we see that the
divergence in $S_{MO}$ should be disregarded since it occurs for a much larger value of $U$
than the divergence in $S_{TO}$. 
These results show that if all the
elements of the susceptibility tensor are considered, RPA calculations are
able to determine the development of new ordered phases that can develop
in multi-orbital systems. Conversely, in multi-orbital systems in which orbital-transverse
order develop, RPA calculations using only the homogeneous susceptibility may lead to 
unphysical results.

\subsection{Strong Coupling Regime}
 
In the regime where the coupling $U$ is sufficiently strong such 
that even in the $s$-model it is energetically favorable 
to locate two electrons with parallel spins at the same
site (and in different orbitals), both the $s$- and $d$-models can be mapped
into effective $t-J-J'$ models and an insulating state with 
magnetic stripes can occur. In
this case the Hubbard repulsion has effectively hybridized both bands causing large
distortions and actually opening a full gap [see Fig.~\ref{akw}(a)]. 
In this strong coupling regime both
models appear to have similar properties, but an insulating magnetic behavior
does not reproduce the experimental behavior observed in several of 
the undoped iron pnictides (such as the 1111 and 122 families). However, this regime could
be applied to the chalcogenides: if $U$ is sufficiently strong the
magnetic behavior that develops in the strong coupling limit is more related
to the hopping parameters and superexchange than to the weak-coupling 
nesting properties of the Fermi surface. While the values of the hopping parameters in
the Hamiltonian are crucial to achieve nesting in weak coupling,\cite{maria} systems in which
nesting is not perfect can develop stripe-like magnetic order if they map into a $t-J-J'$\cite{si}
model in the strong coupling limit such as in the case of the three-orbital model 
for the pnictides.\cite{our3}

\begin{figure}[thbp]
\begin{center}
\includegraphics[width=8.5cm,clip,angle=0]{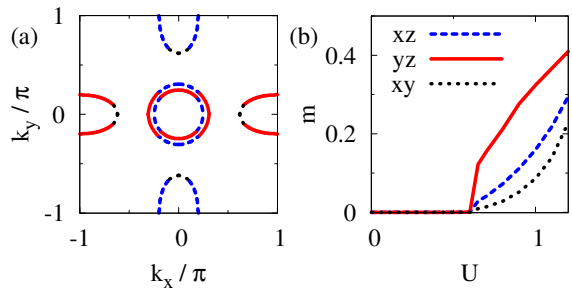}
\caption{(color online) (a) Fermi surface with its orbital composition for the case of a
three-orbital model for the pnictides. (b) Mean-field calculated
orbital-resolved magnetization for the same three-orbital model. The figure was taken from 
Ref.~\onlinecite{our3} for illustration.}
\vskip -0.5cm
\label{3orb}
\end{center}
\end{figure}
 
The results in this section indicate that in the case of the pnictides,
even if the five $d$ orbitals are considered, the $xz$ and $yz$ orbitals are the
most likely to produce the strongest contribution to the metallic stripe magnetic order at weak
or intermediate values of the Hubbard interaction because they are the major constituents of
the FS's with better nesting and because they are degenerate and, thus, there is no  
energy $\Delta$ that needs to be overcome by the interaction. 
This is apparent already in the three-orbital
model for the pnictides, where a mean-field calculation shows that magnetic order 
develops at a finite value of $U$ (see panel (b) of Fig.~\ref{3orb}).\cite{our3} 
In  Fig.~\ref{3orb}(a) it can be observed that the orbital with the best nesting 
associated with
${\bf Q}=(\pi,0)$ is the $yz$ one, indicated by the continuous line. 
A mean-field
calculation of the orbital resolved magnetization $m_{aa}$ for $a$=$xz$, $yz$,
and $xy$, shows that $m_{yz,yz}$ grows very rapidly at the lowest value of $U$.
The magnetizations for the other orbitals develop as $U$ hybridizes and
distorts the original bands. Thus, in the intermediate $U$ regime when magnetism develops,
the $xz$ and $yz$ orbitals are the ones that would develop the stronger magnetization (albeit for different
values of ${\bf Q}$) giving rise to a magnetic metallic phase. Thus, nesting seems to drive the 
magnetization of the $xz/yz$ orbitals while the additional orbital hybridizations that develop due to the 
reconstruction of the FS then drives the smaller magnetization in the remaining orbitals.

\section{Pairing Symmetries}\label{pairing}
 
\begin{figure}[thbp]
\begin{center}
\includegraphics[width=8.0cm,clip,angle=0]{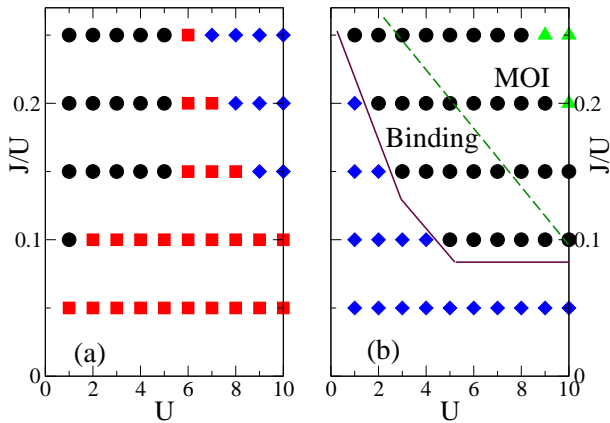}
\caption{(color online) (a) Relative symmetry between the undoped and the
electron-doped ground states for the case of the $d$-orbital model, varying
$J/U$ and $U$. The results were obtained numerically via the Lanczos method using a small
cluster with $N=8$ sites (and following steps already discussed in previous
literature).\cite{Daghofer:2009p1970,nicholson} The circles indicate states with $E_g$ symmetry, squares correspond to $B_{2g}$, and 
diamonds represent $A_{1g}$ symmetric states. (b) Same as (a) but for
the $s$-model with the triangles denoting B$_{1g}$ symmetry. In the region above the continuous
line the two added electrons form a bound state. The
dashed line indicates the boundary for the stability 
of the magnetically ordered insulating (MOI) region in
the undoped state.}
\vskip -0.5cm
\label{sym}
\end{center}
\end{figure}

Regarding the symmetry of the pairing operators corresponding to the models
analyzed here, previous numerical calculations have indicated
a competition between A$_{1g}$, B$_{2g}$, and E$_{g}$ states in the
$d$-model,\cite{moreo,nicholson} as shown in panel (a) of
Fig.~\ref{sym}. The E$_{g}$ states correspond to a $p$-wave spin-triplet state
that becomes destabilized upon the addition of binding-enhancing Heisenberg
terms.\cite{nicholson} The favored pairing operators with the symmetry A$_{1g}$
are all trivial in their orbital composition, i.e. they are intra-orbital with the form
$D^{\dagger}\sigma_0 D$ where
$D^{\dagger}=(d^{\dagger}_{{\bf k},x,\uparrow},d^{\dagger}_{-{\bf k},y,\downarrow})$
in the $({x,y})$ basis, and they remain intraorbital in the $({X,Y})$ basis.
However, the B$_{2g}$ pairing operators have a non-trivial orbital composition given
by $D^{\dagger}\sigma_1 D$ in the basis $({x,y})$, indicating that the pairs
are made of electrons in the two different orbitals. In the $({X,Y})$ basis
the B$_{2g}$ pairing operator becomes
$D'^{\dagger}\sigma_3 D'$ with
$D'^{\dagger}=(d^{\dagger}_{{\bf k},X,\uparrow},d^{\dagger}_{-{\bf k},Y,\downarrow})$.
Thus, in the $(X,Y)$ representation the B$_{2g}$ pairs are intraorbital but
there is an important sign difference between the pairs in the different orbitals which
makes the orbital contribution intraorbital but non trivial. It is interesting to
observe that the intraorbital B$_{1g}$ state found with RPA calculations in the 
five-orbital model for the pnictides \cite{graser} would become interorbital in the
$(X,Y)$ basis.
 
The results for the $s$-model regarding pairing properties are different from 
those
in the $d$-model. Following previous 
investigations,\cite{Daghofer:2009p1970} using the Lanczos method 
we have calculated the relative symmetry 
between the undoped (number of electrons
$N_e=16$) and electron-doped ($N_e=18$) states, as an indicator of the possible
pairing symmetry in the bulk limit. The results are
presented in panel (b) of
Fig.~\ref{sym}, varying $U$ and $J/U$. For small
values of $U$ and $J/U$ the doped ground state has symmetry $A_{1g}$ in agreement
with the $d$-model, although in a different regime of couplings. Increasing $U$
and $J/U$, the $s$-model ground state switches to the $E_{g}$ symmetry, i.e. $p$-wave.
This $p$-symmetry arises from the spatial location of the electrons since the
orbital contribution is trivial.
We have observed that in the small cluster studied here 
the spin-triplet state with ${\bf k}=(0,0)$ 
is almost degenerate with a spin-singlet state with
${\bf k}=(\pi,\pi)$. The possibility of having a spin-singlet
$p$-wave state with wavevector ${\bf k}=(\pi,\pi)$ has been previously discussed 
long ago in the
context of the single-orbital Hubbard model.\cite{scz} In the present case, we need
to remember that here ${\bf k}$ is a pseudo-momentum and in the folded
representation ${\bf k}=(\pi,\pi)$ actually maps into $(0,0)$ so that the actual Cooper
pair, if stabilized, has zero center-of-mass momentum, but the components of the
pair belong to bonding and antibonding bands that could become hybridized 
for the large values of the interactions needed to stabilized these states. 
As indicated in the figure, it
was also found that the $p$-states show binding in the small system studied here.
In addition a small region of bound states with B$_{1g}$ symmetry is found at even larger
couplings. While in the $d$-model our numerical results indicate that the orbital degree of freedom
plays a crucial role in the symmetry of the pairing states, we observe that this is not
the case in the $s$-model. This result seems to indicate that interorbital Cooper pairs are
likely to be present in multi-orbital systems with strongly hybridized bands as it is the case
of the pnictides.

Understanding more deeply why the $s$-model develops its particular pairing properties 
is at this point unnecessary since the model simply provides an illustration of
a system with a similar FS as the $d$-model, and the goal of this 
work was to show that the orbital composition of the
bands plays a crucial role in determining 
the symmetry of the doped states. The examples
that have been discussed here clearly show that models
with the same Fermi surface and the same interactions can have very different 
pairing properties depending on the degree of hybridization of the orbitals. It also 
seems, according to the present results, that the 
relevance of the orbital degree of freedom in determining the pairing 
symmetry is also influenced by the degree of hybridization among those orbitals.

\section{Conclusions}\label{conclusions}
 
Summarizing, numerical and analytical calculations have been performed in 
order to compare the properties of two band models with identical FS's and
interactions, but differing in the degree of hybridization of the orbitals
to form the bands. Despite the nesting properties of the FS's it was discovered
that both models have similar magnetic (insulating) ground states in the 
strong coupling limit, but they are very different in weak and intermediate 
coupling. The $s$-model offers an example in which despite the nesting of the FS and 
the presence of Coulomb interactions, magnetism does not develop in weak coupling. 
However, it was discovered that instead, as a result of the nesting in weak coupling, the
Coulomb interaction stabilizes an orbital-transverse spin ordered state with no local magnetization. This 
state is insulating and is characterized by a gap that could be observed in ARPES
experiments. However, due to the lack of local magnetization, neutron scattering 
experiments would not detect the development of ``generalized spin order''. 
In fact, standard RPA calculations in the $s$-model lead to incorrect results such as 
incommensurate magnetic order in the physical homogeneous channel. However, when
the non-homogeneous components of the susceptibility are taken into account,
RPA reveals the existence of the orbital-transverse spin phase for values of $U$ lower 
than the ones needed to observe the unphysical magnetic state.

It is clear that the physical (homogeneous) magnetic structure factor
depends strongly on the orbital flavor of the bands and for this quantity to 
develop a peak in weak coupling it is necessary that the portions of the FS 
connected by nesting have the same orbital flavor.

The possibility of ``hidden'' magnetic ordering in the pnictides has been proposed by several 
authors\cite{rodriguez,cricchio,bascones}
as an explanation for the unexpectedly low value of the magnetization in several of these 
materials. The hidden order proposed by these authors was ``diagonal'', as the configurations
we presented in Fig.~\ref{cartoon2} after transforming our non-diagonal results into a rotated
orbital basis. However, in multi-orbital systems with more than two orbitals, it may be necessary to
consider the non-diagonal order as well. In theoretical and analytical calculations these non-diagonal
hidden orders are revealed by considering all the components, homogeneous and inhomogeneous, of the
magnetic susceptibility. On the experimental side, ARPES can detect gaps that are opened due to the 
``hidden'' magnetic order but the traditionally used techniques to detect homogeneous magnetic order,
such as neutron scattering, will fail due to the lack of a local magnetization.

We also found indications of quenching of the orbital degree of 
freedom in systems with non-hybridized orbitals. The orbitals do not
appear to play a role in determining the symmetry of the pairing states. This 
degree of freedom, though, is crucial in systems with hybridized orbitals. In 
the case of the pnictides in particular, we have shown that the ground states
with $d$ symmetry found in the literature in models for the pnictides, such
as the B$_{1g}$, can be made interorbital by changing the basis in which the degenerate $xz$ and $yz$ 
orbitals are defined.

The results provided by this work may explain why the end member of the series 
BaFe$_2$(As$_{1-x}$P$_x$)$_2$ is non-superconducting despite displaying the best nesting
of all the compounds in the series.\cite{arnold} If superconductivity necessitates
magnetic fluctuations they may not be sufficiently strong in this compound if there is
no good matching between the flavor of the orbitals in the nested bands.

Finally, our results confirm the perception expressed in the analysis of 
recent photoemission experiments\cite{shimojima} that the
weak coupling nesting mechanism would not be
applicable if indeed a hole-pocket band dominated by the orbital $3z^2-r^2$ (with
no nesting partner in the electron-pocket band)
does develop a robust superconducting gap. Confirming and then understanding
the results of those recent photoemission experiments is indeed very important
for the clarification of several intriguing issues in the challenging physics of the pnictides.
 
\section{Acknowledgments}
  
This work was supported by
the U.S. DOE, Office of Basic Energy Sciences,
Materials Sciences and Engineering Division (A.N., Q.L., W.G., G.M., A.M., E.D.),
by CONICET, Argentina (J.R.),  and by the DFG under the Emmy-Noether program (M.D.).


\begin{thebibliography}{10}
\bibitem{foot} Since the bands are distorted by the interaction 
term of the Hamiltonian, $S_{MB}$ and $S_{TB}$ do
capture the physical properties of the system if 
the interaction is weak enough so that the interacting 
bands resemble the non-interacting ones.
\end{thebibliography}
\end{document}